# Seeing Shapes in Clouds: On the Performance-Cost trade-off for Heterogeneous Infrastructure-as-a-Service

Gordon Inggs, David B. Thomas, George Constantinides
Department of Electrical and Electronic Engineering
Imperial College London
London, United Kingdom
{g.inggs11;d.thomas1;g.constantinides}@imperial.ac.uk

Wayne Luk
Department of Computing
Imperial College London
London, United Kingdom
w.luk@imperial.ac.uk

*Abstract*—In the near future FPGAs will be available by the hour, however this new Infrastructure as a Service (IaaS) usage mode presents both an opportunity and a challenge: The opportunity is that programmers can potentially trade resources for performance on a much larger scale, for much shorter periods of time than before. The challenge is in finding and traversing the trade-off for heterogeneous IaaS that guarantees increased resources result in the greatest possible increased performance. Such a trade-off is Pareto optimal. The Pareto optimal trade-off for clusters of heterogeneous resources can be found by solving multiple, multi-objective optimisation problems, resulting in an optimal allocation of tasks to the available platforms. Solving these optimisation programs can be done using simple heuristic approaches or formal Mixed Integer Linear Programming (MILP) techniques. When pricing 128 financial options using a Monte Carlo algorithm upon a heterogeneous cluster of Multicore CPU, GPU and FPGA platforms, the MILP approach produces a trade-off that is up to 110% faster than a heuristic approach, and over 50% cheaper. These results suggest that high quality performance-resource trade-offs of heterogeneous IaaS are best realised through a formal optimisation approach.

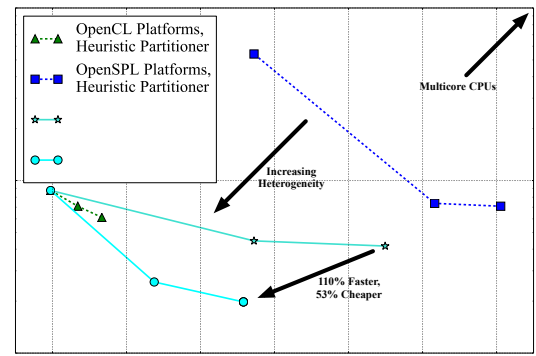

Fig. 1: Latency vs Cost trade-off for 128 option pricing tasks running on 16 heterogeneous Infrastructure-as-a-Service (IaaS) platforms. Full details of the platforms are in Table II.

## I. INTRODUCTION

Heterogeneous clouds are forming. With the use of FPGA-acceleration in a web-based, commodity application [1], as well as the maturation of heterogeneous computing standards, such as OpenCL and OpenSPL; Graphics Processing Units (GPUs) and Field Programmable Gate Arrays (FPGAs) are making inroads in High Performance Computing (HPC) data-centres. As a result, providers are mulling Infrastructure-as-a-Service (IaaS) heterogeneous platforms, and it will soon be possible to make use of diverse heterogeneous accelerators without ever having to own any physical hardware. In this paper, we identify and address a central challenge of this new usage mode: partitioning work within a cluster of heterogeneous computing resources. In doing so, we demonstrate that IaaS FPGAs and GPUs can integrate with and enhance Multicore CPUs in the HPC context.

In the past, HPC programmers targeting heterogeneous platforms were limited by the resources that could be traded for performance. The design space that these programmers inhabited was distorted because any implementations were constrained to the handful of devices that capital resources would allow. Heterogeneous IaaS offers the opportunity to interact with a performance-resource trade-off that seamlessly incorporates both capital and operating costs for a much finer time quantum. As opposed to thinking of using a few devices over a period of years, programmers can now target many more devices for only a few hours.

To realise the opportunity of heterogeneous IaaS, significant implementation challenges have to be overcome. Designs are required that efficiently trade device resource utilisation for improved performance for a wide range of heterogeneous hardware targets [2], [3]. Furthermore, programmers now also have to partition their computational workload across multiple designs running on potentially hundreds of heterogeneous devices. We suggest that this is a partitioning problem, similar to selecting the mapping of subtasks to different architectures or partitioned between software and hardware.

Our initial assumption is that many workloads are composed of multiple, atomic (non-communicating) tasks, as evidenced by the popularity of frameworks such as Pig for Apache Hadoop, and algorithms such as Monte Carlo in computational finance. Furthermore, efficient hardware designs can be realised using heterogeneous computing standards and High Level Synthesis tools. In light of these trends, we propose that the partitioning problem for atomic tasks is best addressed using a formal, multi-objective optimisation approach. The trade-off between performance and resource use is realised by varying the allocation of tasks to platforms. The output of this optimisation process is a Pareto optimal trade-off between the total cost of devices used and a measure of performance achieved. A trade-off that is Pareto optimal allows programmers to achieve greater performance in exchange for a higher cost.

In this paper, we show how to achieve a Pareto optimal cost-performance trade-off for multiple atomic tasks upon multiple, heterogeneous IaaS compute devices. In Figure 1 we illustrate our work with the latency-cost Pareto optimal trade-off for a large computational finance computation of 128 option pricing tasks running on 16 heterogeneous IaaS platforms.

Thus, in this paper we:

1) show how the performance-cost design space for IaaS FPGA resources can be formalised into multiple, multi-objective ILP problems.
2) describe how ILP approaches as well as "common-sense"-based heuristics can be applied to solving these optimisation problems, and so generate the trade-off.
3) evaluate our proposed ILP trade-off generation approach against heuristics using a real workload of 128 financial option pricing tasks upon a heterogeneous cluster incorporating 16 CPU, GPU and FPGA-based Platforms from three major IaaS providers.

Our evaluation shows that a heterogeneous set of platforms can significantly outperform its constituent platforms. Furthermore, adopting an ILP approach to partitioning versus a heuristic one achieves a 110% latency improvement and 50% cost improvement in the best case, and performs no worse in the worst case. As the highly performing partitions achieved using the ILP approach illustrate, HPC datacentres of the future should be heterogeneous, and workload partitioning is best done using a formal optimisation approach.

The section that follows provides a brief review of relevant background material on cloud computing usage models, as well as previous work on workload partitioning in distributed computing contexts. We then describe our proposed approach to the partitioning problem: the necessary resource and performance prediction models; formalising the problem as an ILP; outlining our ILP approach for addressing it as well a heuristic approach. We evaluate the partitioning approaches using a workload of financial option pricing tasks upon a heterogeneous cluster of Multicore CPU, GPU and FPGA-based servers. Finally, we conclude and make recommendations for future work.

## II. BACKGROUND

### A. Cloud Computing Usage Models

Currently there are two dominant utility or "cloud" computing models: application services, and IaaS [2]–[4]. In the application mode, users pay for access to a service, such as Gmail or SAP, that is provided using computing resources that are hosted in a datacentre owned by the service provider, or that the service provider has leased from a IaaS provider.

TABLE I: Comparison of IaaS offerings. Providers are Microsoft Azure (*MA*), Google Compute Engine (*GCE*) and Amazon Web Services (*AWS*). Prices as of April, 2015.

| Provider | Instance Type | Instance Name | Time Quantum (minutes) | Theoretical Peak Performance (GFLOPS) | Rate ($/hour) |
|---|---|---|---|---|---|
| MA | CPU | A4 | 1 | 416 | 0.592 |
| GCE | CPU | n1-highcpu-8 | 10 | ≈400 | 0.352 |
| AWS | CPU | c3.4xlarge | 60 | 883 | 0.924 |
| AWS | GPU | g2.2xlarge | 60 | 2289 | 0.650 |

IaaS providers, such as Amazon Web Services (AWS), Google Compute Engine (GCE) or Microsoft Azure (MA), allow for compute resources to be leased directly. These resources are abstracted as virtual servers or instances, accessed via the Internet using protocols such as SSH, which the user may then configure with the desired software. Resources are priced using a rate quoted on a per instance type, per time increment or quantum basis. This rate reflects both the operating and capital expenses of the resources of that instance for the time quantum as well as the provider's profit margin.

The key characteristics of some of the instances from the most popular IaaS providers are reported in Table I. A key feature of each offering is the length of the time quantum, i.e. the minimum increments of time for which the user will be charged. We also observe that the rate reflects performance within the CPU category, hence an instance with twice the peak compute capability of another will roughly cost twice as much.

However, a further observation is that *between* heterogeneous device categories, such as CPUs and GPUs, the link between pricing and performance does not hold. For example the AWS GPU instance listed theoretically offers an extremely attractive performance to cost ratio relative to traditional CPUs, but is priced in the middle of the CPU price range.

### B. Workload Partitioning for Heterogeneous Computing

The problem of partitioning computational tasks across distributed, heterogeneous computing resources has been widely studied. The general scenario often considered in the literature is a set of atomic tasks being partitioned across multiple platforms of different capabilities [5], [6]. In this scenario, it is assumed that if a task is allocated to a resource, it will fully occupy that resource until completed. The partitioning is also performed statically, in advance of the tasks' execution, using estimates of performance metrics.

More recent work has considered dynamic allocation during task runtime [2], [3], however this effectively takes the form of static allocation performed on a regular interval with updated task information.

In the atomic task allocation scenario, the general objective is to optimise a measure of performance, often the workload latency or makespan. The *makespan* is the latency between when the first task is initiated until the last result returned. As the tasks are being evaluated on multiple platforms, the makespan is equivalent to the latency of the platform that takes the longest to complete its assigned tasks.

Two of the suggested approaches in the literature to the performance optimisation problems: *Naive Heuristics* [2], [5] - a simple algorithm is applied to allocate tasks to the available resources. The quality of the partitions produced are highly dependent on the particular tasks and platforms concerned. *Integer Linear Programming* [6] - the partitioning problem is formulated as an optimisation program which can then be solved using ILP techniques, such as the branch and bound algorithms as well as multiple heuristics.

Generally heuristic approaches have been the most studied. Braun's comprehensive study [5] found that simpler heuristics achieve better results than more complex ones. We suggest that this indicates that the truly optimal approach is case-specific, dependent upon subtle dynamics between the task and platforms concerned. ILP appears to be an understudied approach, usually applied only in environments of pressing resource constraint [6]. This lack of attention is likely due to the NP-hard complexity of ILPs in general, and NP-complete in the binary case, prompting concerns over the uncertainty of the time spent finding a solution.

## III. OUR PARTITIONING APPROACH

In this section we describe our approach to partitioning workloads of atomic tasks across heterogeneous IaaS resources so as to achieve a Pareto optimal trade-off between resource use and performance. Throughout our explanation, we use the example of Monte Carlo algorithm-based, financial option pricing tasks.

### A. Latency and Cost Models

As described in the previous section, partitioning approaches require some estimation of the critical task characteristics, such as the makespan and financial cost. Hence, models of these characteristics have to be used to predict the performance of the available implementations.

The latency and cost models that we use in Monte Carlo option pricing tasks are given in Equation 1.

$$L(N) = \beta N + \gamma \quad (1a)$$

$$C(L(N)) = \left\lceil \frac{L(N)}{\rho} \right\rceil \pi \quad (1b)$$

The latency model given in Equation 1a is a linear one, comprised of proportional ($\beta N$) and constant ($\gamma$) terms. The constant term reflects the overhead in initiating the task on a platform, incorporating time spent in communication, device configuration in the FPGA case, etc. The proportional term grows with the input variable ($N$), reflecting the growth in the number of operations as the task increases in scale. This model reflects the $O(N)$ complexity of the Monte Carlo algorithm, and would need additional polynomial terms for tasks that are more computationally complex.

To find the values of the latency model coefficients ($\beta$ and $\gamma$), we propose a benchmarking procedure for all of the tasks upon all of the available target devices, using a set of $N$ and latency values, as well as weighted least squares regression to solve for the model parameters, $\beta$ and $\gamma$.

The cost model given in Equation 1b reflects the IaaS model described in the previous section. The task latency is divided by the time quantum ($\rho$), which is then rounded up. This is then multiplied by the platform rate ($\pi$). As this model is expressed in terms of the latency model, it is easily generalised, provided an appropriate latency model is available.

Equation 2 describes how we suggest finding the rate ($\pi$) for IaaS FPGAs in the current absence of observable market prices.

$$\begin{aligned} \pi &= \text{DBR} \times \text{RDP} \\ \text{DBR} &= (\text{TCO} + \text{PM})\frac{\rho}{P} \end{aligned} \quad (2)$$

The rate is given by the Device Base Rate (DBR), which is the cost per device in the datacentre for the specified time quantum ($\rho$), scaled by the Relative Device Performance (RDP), the performance of the device relative to the other devices *of the same type* in the datacentre, as per the precedent observed in the market currently. The DBR is given by the annual total cost of ownership for that device (TCO) plus the profit margin (PM) scaled by the time quantum ($\rho$) to year ($P$) ratio, i.e. $\frac{\rho}{P}$. To find the TCO, we suggest using a total cost of ownership model for datacentres, such as the simple model published by the Uptime Institute [7].

### B. Latency Minimisation on a Budget

The models in the previous subsection describe a single Monte Carlo task upon a single platform. In this subsection we show how these models can be used to trade between characteristics for a workload of $\tau$ tasks upon a cluster of $\mu$ platforms.

A task-platform allocation could be binary, of whole tasks to platforms. However this doesn't take advantage that tasks are often composed of parallel subtasks. If the degree of parallelism, such as $N$ in the Monte Carlo case, in the set of tasks is sufficiently large, then such an allocation can be real-valued between 0 and 1, representing the proportions of tasks allocated to different platforms. By allowing this *relaxed* allocation, we can cast the partitioning problem as a financial cost constrained, Mixed ILP makespan optimisation problem. Equation 3 gives our formulation of this problem, with the cost constraint ($C_k$) and task-platform allocation ($\boldsymbol{A}$).

$$\begin{aligned} \underset{\boldsymbol{A} \in \mathbb{R}_+^{\mu \times \tau}}{\text{minimise}} \quad & F_L(\vec{G}_L(\boldsymbol{A})) \\ \text{subject to} \quad & \sum_{i=1}^{\mu} A_{i,j} = 1 \quad j = 1, 2, \ldots, \tau \\ & F_C(\vec{G}_C(\boldsymbol{A})) \leq C_k \quad C_k \in \mathbb{R}_+ \end{aligned} \quad (3)$$

where:

$$\begin{aligned} F_L(\vec{G}_L(\boldsymbol{A})) &= \max(\vec{G}_L(\boldsymbol{A})) \\ \vec{G}_L(\boldsymbol{A}) &= ((\boldsymbol{\beta} \circ \boldsymbol{N}) \circ \boldsymbol{A} + \boldsymbol{\gamma} \circ \lceil \boldsymbol{A} \rceil) \cdot \boldsymbol{1} \\ & \boldsymbol{\beta}, \boldsymbol{\gamma} \in \mathbb{R}_+^{\tau \times \mu}, \boldsymbol{N} \in \mathbb{Z}_+^{\tau \times \mu} \\ F_C(\vec{G}_C(\boldsymbol{A})) &= \boldsymbol{1}^T \cdot \vec{G}_C(\boldsymbol{A}) \\ \vec{G}_C(\boldsymbol{A}) &= \vec{\pi} \circ \left\lceil \frac{\vec{G}_L(\boldsymbol{A})}{\rho} \right\rceil \quad \vec{\pi} \in \mathbb{R}_+^{\mu} \end{aligned}$$

The Latency and Cost models for the financial option pricing tasks that were given in Equation 1 are captured in what we define as the task reduction functions, $\vec{G}_L(\boldsymbol{A})$ and $\vec{G}_C(\boldsymbol{A})$, which provide the platform latency and cost for a given allocation ($\boldsymbol{A}$). In both task reduction functions, $\circ$ represents the Hadamard or entrywise product of matrices or vectors.

In what we define the platform reduction functions, $F_L(\vec{G}_L(\boldsymbol{A}))$ and $F_C(\vec{G}_C(\boldsymbol{A}))$, the platforms' characteristics are combined to a scalar value. In the latency case, this is the makespan, while for financial cost this is the total IaaS utilisation cost.

Many formal optimisation frameworks such as SCIP [8] accept problems in the form given in equation 3, however they do not support non-linear objective or constraint functions such as the maximum and ceiling functions used in the platform latency ($F_L(\vec{G}_L(\boldsymbol{A}))$) and cost reduction ($\vec{G}_C(\boldsymbol{A})$) functions. We now show how these non-linear functions can be captured in equation 4.

$$\begin{aligned}
\underset{\boldsymbol{A} \in \mathbb{R}_+^{\mu \times \tau}}{\text{minimise}} \quad & F_L \\
\text{subject to} \quad & \sum_{i=1}^{\mu} A_{i,j} = 1 \quad j = 1, 2, \ldots, \tau \\
& \vec{G}_L(\boldsymbol{A}) \leq F_L \\
& A_{i,j} \leq B_{i,j} \\
& \boldsymbol{B} \in \{0,1\}^{\mu \times \tau}, i = 1, 2, \ldots, \mu, j = 1, 2, \ldots, \tau \\
& \frac{\vec{G}_{L,i}(\boldsymbol{A})}{\rho_i} \leq D_i \quad \vec{\rho} \in \mathbb{Z}_+^{\mu}, \vec{D} \in \mathbb{Z}_+^{\mu}, i = 1, 2, \ldots, \mu \\
& F_C(\vec{D}) \leq C_k \quad C_k \in \mathbb{R}_+
\end{aligned} \quad (4)$$

where:
$$\vec{G}_L(\boldsymbol{A}) = ((\boldsymbol{\beta} \circ \boldsymbol{N}) \circ \boldsymbol{A} + \boldsymbol{\gamma} \circ \boldsymbol{B}) \cdot \boldsymbol{1}$$
$$\boldsymbol{\beta}, \boldsymbol{\gamma} \in \mathbb{R}_+^{\tau \times \mu}, \boldsymbol{N} \in \mathbb{Z}_+^{\tau \times \mu}$$
$$F_C(\vec{D}) = \vec{D}^T \cdot \vec{\pi} \quad \vec{\pi} \in \mathbb{R}_+^{\mu}$$

We have transformed the non-linear functions in the partitioning problem into additional dependent variables and constraints. Firstly, an additional real variable (FL) is introduced that is constrained to being greater than all of the individual platform latencies, capturing the maximum function in platform reduction function ($F_L(\boldsymbol{A})$). A binary variable ($\boldsymbol{B}$) greater than or equal to the allocation variable, captures the ceiling function in the latency task reduction function ($\vec{G}_L(\boldsymbol{A})$). Finally an integer variable ($\boldsymbol{D}$) captures the ceiling function in the cost reduction function ($\vec{G}_C(\boldsymbol{A})$).

### C. Finding the Latency-Cost Tradeoff

The previous subsection describes how to minimise latency for a single, fixed cost constraint, however we seek a method for finding the resource-performance trade-off. The previously described program can be used to find such a trade-off by using ILP evaluation tools such as SCIP [8], through the $\epsilon$-constraint method as described by Kirlik et al [9]. By contrast, we also describe a heuristic approach to finding different resource-performance trade-off points.

For our example of a latency-cost trade-off, the same procedure for both the ILP and heuristic approaches is given below.

*1) Find the upper cost bound ($C_U$):* For the ILP approach this can be found by minimising the latency without the cost constraint, i.e. $F_C(\vec{D}) \leq C_k$, as this will give the maximum cost on the Pareto curve. Heuristically, this can be found by dividing work inversely proportional to the individual makespans of the available platforms.

*2) Find the lower cost bound ($C_L$):* For both the ILP and heuristic approaches, the lowest cost possible is found by allocating all the tasks to the single platform that completes all of the tasks as cheaply as possible. This gives the lowest cost on the Pareto curve.

*3) Iterate between $C_L$ and $C_U$:* For ILP, as per $\epsilon$-constraint method, run the program outlined in Equation 4 for a set of cost constraints ($C_k$) spaced evenly between the upper and lower bounds, for the desired degree of granularity. For the heuristic approach, a linear combination of the normalised latency-cost product can be used each platform. As the weighting of the cost is increased, the trade-off should move from $C_U$ to $C_L$.

## IV. EVALUATION

We now evaluate the claims that we have made with regards to modelling task-device latency and cost characteristics in advance, the efficiency of different partitioning approaches and finally, the generation of a Pareto trade off for cost and performance.

### A. Experimental Setup

*1) Tasks:* The computational workload that we used is the pricing of 128 financial option pricing tasks using the Monte Carlo algorithm. The algorithm is compute bound, with random generation accounting for the bulk of the computations. In addition to all of the option pricing tasks being independent, the simulations within each task can be computed in parallel, hence these can be split between multiple platforms. The fixed parameters for the pricing task operations were generated from within the values from the Kaiserslautern option pricing benchmark[1]. The number of simulations per Monte Carlo task ($\boldsymbol{N}$) was set so as to achieve an accuracy of \$0.001 for each task.

*2) Platforms:* Table II provides the details of the heterogeneous cluster that we have used. The cluster is largely made up of Maxeler and Altera FPGA accelerator boards that communicate with the host using PCIe. The FPGA platforms were programmed using both the OpenSPL and OpenCL heterogeneous computing standards, and the Maxeler and Altera High Level Synthesis tools. The two CPUs are those provided by MA and GCE, and are programmed using POSIX and GCC, while the GPU is provided by AWS and programmed using OpenCL and the Nvidia SDK. The rate for the FPGA devices was calculated using Equation 2, with the parameters given in Table III, and the RDP weighting for each device calculated using the relative application performance.

---
[1] http://www.uni-kl.de/en/benchmarking/option-pricing/

The heterogeneous standards deliver portable performance: the same OpenSPL designs delivers similar performance upon the two platform targeted, despite being implemented on FPGAs from different vendors. Similarly, the difference in performance between the OpenCL GPU and FPGA implementations can be explained almost entirely by the difference in clock rate, suggesting performance portability across device architectures.

*3) Software Framework:* For task implementation, execution and partitioning, we used the Forward Financial Framework[2]($F^3$), an Open Source, Python-based Financial Application Framework. $F^3$ allows for financial problems to be expressed using a library of domain specific objects. The problems can then be evaluated on range of distributed, heterogeneous platforms efficiently [10].

To support the partitioning of tasks, we have extended $F^3$ to partition workloads using the approaches in Section III.C. To support the ILP approach, we used SCIP [8] as a black-box Mixed ILP optimiser, with Equation 4 as the input program.

*B. Method*

First we verified the models that we used as inputs into our partitioning approaches. To verify the cost model, we applied the same cost methodology to the IaaS offerings from Amazon as well as a hypothetical FPGA datacentre. For the latency model, we measured the relative error of the latency predictions for 10 minutes of benchmarking. We used heuristic and ILP approaches to finding partitions for our computational workload for multiple budgets, including the lower and upper cost bound. Finally, we used the partitioning approaches to generate latency and cost curves using the model data as inputs. We then ran the resulting partitions on our experimental hardware that make up the curve, verifying the validity of the partitioner outputs.

*C. Results and Discussion*

*1) Cost and Latency Models:* Our latency model is verified in Figure 2. The relative error of the latency predicted versus that seen in reality is within 10% for problems many times the size of the benchmarking subset used. As we will show below, this is sufficiently accurate to generate a workload partition.

We have verified our cost model in Table III. We used the Uptime Institute's datacentre cost model updated to 2015 prices, applied to create hypothetical CPU[3], GPU[4] and FPGA[5] IaaS offerings. We have compared the CPU and GPU to AWS's IaaS offering.

The relatively lower capital recovery periods we used for CPUs and GPUs reflect the faster development cycle of these devices as well as the competitive IaaS market. The number of devices given is the number that would fit within the standard datacentre in the Uptime Institute's model.

Both the GPU and CPU rates are very close to those observed in reality, however both are several percent below those seen in the market. This is most likely due to an underestimation of the operating costs of the datacentre.

---

[2]https://github.com/Gordonei/ForwardFinancialFramework
[3]Full CPU model: http://bit.ly/1IdJgNg
[4]Full GPU model: http://bit.ly/1GbKVlT
[5]Full FPGA mode: http://bit.ly/1MKjGmc
[6]http://aws.amazon.com/ec2/pricing/

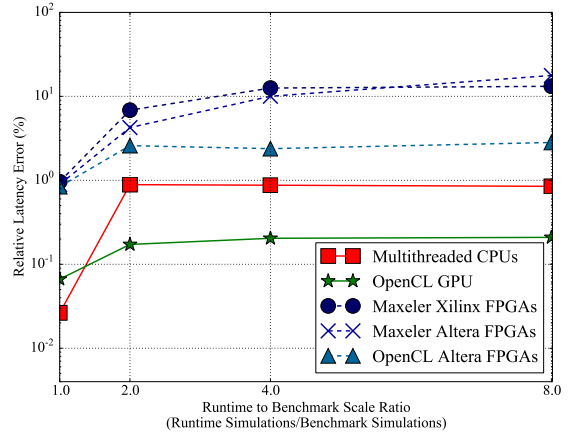

Fig. 2: Latency model prediction error characterisation

TABLE III: Cost model applied to CPUs, GPUs and FPGAs

| Parameter | FPGA Model | GPU Model | CPU Model |
|---|---|---|---|
| Device Capital Cost | $5370 | $3120 | $2530 |
| Energy Use | 50W | 135W | 115W |
| Number of Devices | 5181 | 5181 | 5181 |
| Capital Recovery Period | 5 years | 2 years | 2 years |
| Charged Usage | 80% | 80% | 90% |
| Profit Margin | 20% | 20% | 20% |
| Calculated Device Rate | $0.46/hour | $0.64/hour | $0.50/hour |
| Observed Device Rate[6] | - | $0.65/hour | $0.53/hour |

*2) Partition Generation:* In Table IV, the latency is given for three cost constraints, using the two different partitioning approaches. Both share the same lower cost bound, which is to allocate all of the work to the GPU platform.

The ILP approach demonstrates a significant improvement over the heuristic in the median and upper cost bound values. This difference is explained by the heuristic approach only considering absolute latency and cost, and not taking into account the non-linearities in the latency due to the constant setup time, and in the cost due to the length of time quantum. A good example of this is the CPU platforms, which the heuristic approach does not consider at all, but the ILP does, due to the reduced time quanta both offer.

*3) Trade-off Comparison:* In Figure 3, we plot the latency-cost design spaces for the two partitioning approaches. For each approach we plot the model data latency-cost trade-off

TABLE IV: Latency-Cost Trade-off for Heuristic and Integer Linear Programming (ILP) Approaches

| Cost Level | Metric | Heuristic | ILP | $\frac{\text{Heuristic}}{\text{ILP}}$ |
|---|---|---|---|---|
| Cheapest ($C_L$) | Cost ($) | 1.950 | 1.950 | 1.0 |
| | Latency (S) | 8760.420 | 8760.420 | 1.0 |
| Median ($C_k$) | Cost ($) | 7.445 | 4.749 | 1.57 |
| | Latency (S) | 4468.920 | 2582.483 | 1.73 |
| Fastest ($C_U$) | Cost ($) | 10.990 | 7.160 | 1.53 |
| | Latency (S) | 4172.144 | 1979.448 | 2.11 |

TABLE II: Experimental Heterogeneous Computing Platforms. IaaS providers are Microsoft Azure (*MA*), Google Compute Engine (*GCE*) and Amazon Web Services (*AWS*). Performance was measured using the Kaiserslautern option pricing benchmark.

| # | Provider | Device | Programming Standard (Tool) | Lookup Tables | Flipflops | BRAMS | DSPs | Clockrate (Ghz) | Application Performance (GFLOPS) | Rate ($/hour) |
|---|---|---|---|---|---|---|---|---|---|---|
| 4 | - | Xilinx Virtex 6 475T | OpenSPL (MaxCompiler 2013.2.2) | 298k | 595k | 1064 | 2016 | 0.2 | 111.978 | 0.438 |
| 8 | - | Altera Stratix V GSD8 | OpenSPL (MaxCompiler 2013.2.2) | 695k | 1050k | 2567 | 3926 | 0.18 | 112.949 | 0.442 |
| 1 | - | Altera Stratix V GSD5 | OpenCL (Altera SDK 14.0) | 457k | 690k | 2014 | 3180 | 0.25 | 176.871 | 0.692 |
| 1 | AWS | Nvidia Grid GK104 | OpenCL (Nvidia SDK 6.0) | - | - | - | - | 0.8 | 556.085 | 0.650 |
| 1 | MA | Intel Xeon E5-2660 | POSIX (GCC 4.8) | - | - | - | - | 2.2 | 4.160 | 0.480 |
| 1 | GCE | Intel Xeon | POSIX (GCC 4.8) | - | - | - | - | 2.0 | 6.022 | 0.352 |

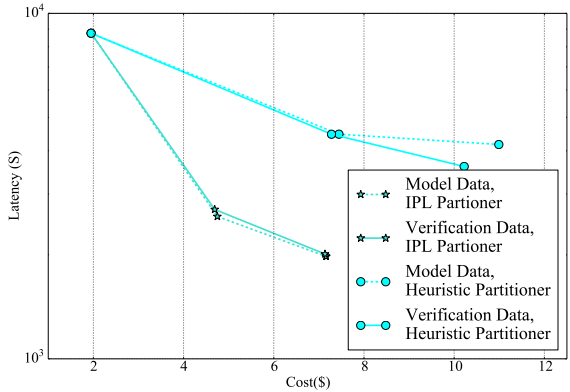

Fig. 3: Partitioner performance model predictions vs measured.

versus the actual trade-off realised when we ran the partitions.

Both approaches' model curves are sufficiently close to the actual data trade-off that a programmer could use these approaches to balance their objectives in advance of actual problem execution. A notable outlier is the upper cost bound of the heuristic approach, where that seen in reality is 12% quicker and 7% cheaper than what is projected by the model. This is consistent with the 10% mean error seen in the latency prediction models.

## V. CONCLUSION

In this paper we addressed the challenge of partitioning workloads across heterogeneous IaaS resources so that lower latencies can be achieved at increased cost. We showed that predictive runtime characteristic models combined with a multi-objective optimisation approach provide an effective methodology for generating Pareto optimal performance-cost trade-offs. We also evaluated two distinct methods for partitioning, showing that our proposed Mixed ILP approach yields a more efficient design spaces than a heuristic one.

Furthemore, our work helps makes the case for heterogeneous IaaS, demonstrating significant performance improvement and cost saving through heterogeneous architectures compared to just using conventional CPUs. However, we argue that these benefits are only realisable if programmers have a means to balance their objectives efficiently, such as our approach to workload partitioning.

In the future we would like to increase the scale of these experiments, both in terms of the number of platforms and tasks, as well as in terms of range of data points explored. There is also significant scope for tuning the partitioners utilised.

## ACKNOWLEDGEMENTS

We are grateful for the funding support from the Oppenheimer Memorial Trust as well as the South African National Research Foundation. We would also like to acknowledge the resources obtained through the Maxeler, Altera and Nallatech University Programs as well an Amazon Web Services Educational grant.

## REFERENCES

[1] A. Putnam, A. M. Caulfield, E. S. Chung, D. Chiou, K. Constantinides, J. Demme, H. Esmaeilzadeh, J. Fowers, G. P. Gopal, J. Gray, M. Haselman, S. Hauck, S. Heil, A. Hormati, J. Y. Kim, S. Lanka, J. Larus, E. Peterson, S. Pope, A. Smith, J. Thong, P. Y. Xiao, and D. Burger, "A reconfigurable fabric for accelerating large-scale datacenter services," in *Proceedings of International Symposium on Computer Architecture*, 2014, pp. 13–24.

[2] E. Oneill, J. McGlone, P. Milligan, and P. Kilpatrick, "SHEPARD: Scheduling on heterogeneous platforms using application resource demands," in *Proceedings of 2014 22nd Euromicro International Conference on Parallel, Distributed, and Network-Based Processing*. IEEE, Feb. 2014, pp. 213–217.

[3] J. G. F. Coutinho, O. Pell, E. O'Neill, P. Sanders, J. McGlone, P. Grigoras, W. Luk, and C. Ragusa, "HARNESS project: Managing heterogeneous computing resources for a cloud platform," pp. 324–329, 2014.

[4] J. A. Varia and S. A. Mathew, "Overview of AWS," Amazon, Tech. Rep., 2014.

[5] T. Braun, H. J. Siegel, N. Beck, L. Bölöni, M. Muthucumaru, A. Reuther, J. P. Robertson, M. D. Theys, B. Yao, D. Hensgen, and R. F. Freund, "A Comparison of Eleven Static Heuristics for Mapping a Class of Independent Tasks onto Heterogeneous Distributed Computing Systems," *J. Parallel Distrib. Comput.*, vol. 61, no. 6, pp. 810–837, Jun. 2001.

[6] N. Fisher, J. Anderson, and S. Baruah, "Task Partitioning upon Memory-Constrained Multiprocessors," *11th IEEE Int. Conf. Embed. Real-Time Comput. Syst. Appl.*, pp. 416–421, 2005.

[7] J. Koomey, K. Brill, P. Turner, J. Stanley, and B. Taylor, "A Simple Model for Determining True Total Cost of Ownership for Data Centers," Tech. Rep., 2008.

[8] T. Achterberg, "Scip: Solving constraint integer programs," *Mathematical Programming Computation*, vol. 1, no. 1, pp. 1–41, 2009.

[9] G. Kirlik and S. Sayın, "A new algorithm for generating all non-dominated solutions of multiobjective discrete optimization problems," *European Journal of Operational Research*, vol. 232, no. 3, pp. 479–488, 2014.

[10] G. Inggs, D. Thomas, and W. Luk, "A Domain Specific Approach to Heterogeneous Computing: From Availability to Accessibility," in *Proc. First Int. Work. FPGAs Softw. Program. (FSP 2014)*, Aug. 2014.